\documentclass[11pt,a4paper]{article}

\usepackage[T1]{fontenc}
\usepackage[utf8]{inputenc}

\usepackage[margin=2.5cm]{geometry}
\usepackage{amsmath,amssymb}
\usepackage{graphicx}
\usepackage{booktabs}
\usepackage{tabularx}
\usepackage{multirow}
\usepackage{array}
\usepackage{xcolor}
\usepackage{enumitem}
\usepackage{hyperref}
\usepackage{url}
\usepackage{cite}
\usepackage{float}
\usepackage{caption}
\usepackage{subcaption}
\usepackage{listings}
\usepackage{fancyhdr}
\usepackage{abstract}
\usepackage{parskip}
\usepackage{titlesec}
\usepackage{tcolorbox}

\definecolor{codeblue}{RGB}{30,80,160}
\definecolor{codegray}{RGB}{128,128,128}
\definecolor{codegreen}{RGB}{0,128,64}
\definecolor{codered}{RGB}{180,30,30}
\definecolor{rowgray}{RGB}{245,245,250}
\definecolor{headerblue}{RGB}{26,26,46}
\definecolor{accentblue}{RGB}{74,158,255}

\hypersetup{
  colorlinks=true,
  linkcolor=codeblue,
  citecolor=codeblue,
  urlcolor=codeblue,
  pdftitle={Quantum-Safe Code Auditor: An LLM-Augmented Static Analysis Tool
            for Detecting and Prioritising Post-Quantum Cryptography Migration Risks},
  pdfauthor={Animesh Shaw},
  pdfsubject={Cryptography, Security, Post-Quantum},
  pdfkeywords={post-quantum cryptography, static analysis, LLM, variational quantum eigensolver,
               NIST, migration, vulnerability detection}
}

\titleformat{\section}{\large\bfseries\color{headerblue}}{{\thesection.}}{0.5em}{}
\titleformat{\subsection}{\normalsize\bfseries\color{headerblue}}{{\thesubsection}}{0.5em}{}

\tcbuselibrary{skins,breakable}
\newtcolorbox{keybox}[1]{
  colback=rowgray, colframe=accentblue!70,
  fonttitle=\bfseries\small, title=#1,
  left=4pt, right=4pt, top=4pt, bottom=4pt
}

\begin{document}

\title{\textbf{Quantum-Safe Code Auditing: LLM-Assisted Static\\ 
Analysis and Quantum-Aware Risk Scoring for\\ Post-Quantum Cryptography Migration}}

\author{
  Animesh Shaw\\
  \texttt{animesh15b@iimk.edu.in}\\[0.3em]
  \small{Independent Researcher}
}

\date{}

\maketitle
\thispagestyle{fancy}

\begin{abstract}
\noindent
The impending arrival of cryptographically relevant quantum computers (CRQCs) threatens the security foundations of modern software: Shor's algorithm breaks RSA, ECDSA, ECDH, and Diffie-Hellman, while Grover's algorithm reduces the effective security of symmetric and hash-based schemes. Despite NIST standardising post-quantum cryptography (PQC) in 2024 (FIPS~203 ML-KEM, FIPS~204 ML-DSA, FIPS~205 SLH-DSA), most codebases lack automated tooling to inventory classical cryptographic usage and prioritise migration based on quantum risk. We present \emph{Quantum-Safe Code Auditor}, a quantum-aware static analysis framework that combines (i) regex-based detection of 15 classes of quantum-vulnerable primitives, (ii) LLM-assisted contextual enrichment to classify usage and severity, and (iii) risk scoring via a Variational Quantum Eigensolver (VQE) model implemented in Qiskit~2.x, incorporating qubit-cost estimates to prioritise findings. We evaluate the system across five open-source libraries—python-rsa, python-ecdsa, python-jose, node-jsonwebtoken, and Bouncy Castle Java—covering 5{,}775 findings. On a stratified sample of 602 labelled instances, we achieve 71.98\% precision, 100\% recall, and an F1 score of 83.71\%. All code, data, and reproduction scripts are released as open-source~\cite{github2026}.
\end{abstract}

\vspace{0.6em}

\section{Introduction}
\label{sec:intro}

The security of public-key cryptography deployed in virtually all production software
today rests on the computational hardness of integer factorisation and the discrete
logarithm problem.
Shor's quantum algorithm~\cite{shor1994} solves both in polynomial time.
Gidney and Ekerå~\cite{gidney2021} showed that RSA-2048 can be broken using approximately
4{,}096 logical qubits---well within the envelope of machines forecast for the early to
mid-2030s.
Elliptic-curve variants (ECDSA, ECDH, X25519, Ed25519) require fewer qubits still:
roughly 2{,}330 for NIST P-256~\cite{roetteler2017}.
Grover's algorithm~\cite{grover1996} provides a quadratic speedup over brute-force search,
effectively halving the bit-security of symmetric ciphers and hash functions.
AES-128 drops to 64-bit security; SHA-256 to 128-bit; legacy primitives such as
3DES, RC4, MD5, and SHA-1 that are already marginal in the classical setting become
untenable.

The threat is not purely future-facing.
The \emph{Harvest Now, Decrypt Later} (HNDL) strategy allows adversaries to intercept
and archive encrypted traffic today, decrypting it retrospectively once a CRQC is
operational.
The NSA's Commercial National Security Algorithm Suite 2.0
(CNSA~2.0)~\cite{cnsa2} mandates full PQC migration by 2030, and the August 2024
publication of NIST FIPS 203~\cite{nist203}, 204~\cite{nist204}, and 205~\cite{nist205}
provides concrete, standardised replacements.

Yet the practical challenge of migration remains unaddressed at the tooling level.
Most development teams lack visibility into \emph{where} classical cryptography lives
in their codebases, let alone a ranked list of which usages carry the greatest quantum
risk.
Existing static-analysis tools (CryptoGuard~\cite{cryptoguard}, Bandit, SonarQube) detect
cryptographic misuse in the classical sense---weak key sizes, deprecated APIs---but do not
model quantum cost, do not integrate VQE-based threat estimation, and do not provide
NIST PQC migration guidance as output.

\paragraph{Contributions} ~\\[0.2em]
This paper makes the following contributions:\\[0.1em]
\begin{enumerate}[leftmargin=*, nosep]
  \item A \textbf{three-tier pipeline} combining regex scanning, LLM enrichment, and VQE
        threat scoring for end-to-end quantum cryptography risk assessment
        (Section~\ref{sec:architecture}).
  \item A \textbf{VQE threat model} that translates cryptographic algorithm properties
        into a continuous 0--10 risk score using a Qiskit 2.x implementation, providing an
        intuitive prioritisation signal for engineering teams
        (Section~\ref{sec:implementation}).
  \item An \textbf{evaluation across five open-source repositories} (602 labelled findings,
        5{,}775 total) with per-algorithm and per-tier precision, recall, and F1 scores
        (Section~\ref{sec:evaluation}).
  \item An \textbf{open-source release} of all tool code, evaluation data,
        labelled samples, and reproduction scripts.
\end{enumerate}

\section{Background and Related Work}
\label{sec:background}

\subsection{Quantum Algorithms and Cryptographic Impact}

\textbf{Shor's algorithm}~\cite{shor1994} (1994) factors an $n$-bit integer in
$O(n^3)$ quantum gate operations.
Its practical implication is the complete break of every public-key system whose security
relies on RSA, DSA, or elliptic-curve groups: ECDSA, ECDH, X25519, Ed25519, and
Diffie-Hellman are all vulnerable.
Resource estimates have tightened over time; Gidney and Ekerå~\cite{gidney2021}
demonstrated that RSA-2048 requires $\approx$4{,}096 logical qubits, far fewer than
earlier analyses.

\textbf{Grover's algorithm}~\cite{grover1996} (1996) provides an unstructured search
speedup from $O(N)$ to $O(\sqrt{N})$.
For symmetric key search this halves effective security: AES-128 offers $2^{64}$
post-quantum security rather than $2^{128}$; SHA-256 drops to 128-bit pre-image
resistance.
NIST SP 800-131A recommends key lengths of $\geq$256 bits for symmetric schemes in
the post-quantum era.

\subsection{NIST PQC Standards}

After a multi-year competition, NIST published three final PQC standards in August 2024:
\begin{itemize}[nosep]
  \item \textbf{FIPS 203}~\cite{nist203}: ML-KEM (Module Lattice Key Encapsulation
        Mechanism, based on Kyber).
        Provides IND-CCA2 secure key encapsulation as a direct replacement for RSA and
        ECDH key exchange.
  \item \textbf{FIPS 204}~\cite{nist204}: ML-DSA (Module Lattice Digital Signature
        Algorithm, based on Dilithium).
        Provides EUF-CMA secure signatures as a replacement for ECDSA and RSA signatures.
  \item \textbf{FIPS 205}~\cite{nist205}: SLH-DSA (Stateless Hash-Based Digital Signature
        Algorithm, based on SPHINCS+).
        Provides a hash-based fallback signature scheme with conservative security
        assumptions independent of lattice hardness.
\end{itemize}

NSA CNSA~2.0~\cite{cnsa2} sets 2030 as the mandatory migration deadline for National
Security Systems (NSS), providing the temporal urgency that motivates automated migration
tooling.

\subsection{Harvest-Now Decrypt-Later}

Mosca~\cite{mosca2018} formalised the HNDL risk as follows: if an adversary intercepts
encrypted data today and a CRQC becomes operational before the data loses sensitivity,
the data is compromised.
The migration timeline therefore begins \emph{now}, not at CRQC availability.

\subsection{Related Tooling}

\textbf{CryptoGuard}~\cite{cryptoguard} (CCS 2019) performs backward dataflow analysis to
detect 22 classes of classical cryptographic misuse in Java.
It targets incorrect API usage (e.g., ECB mode, short keys) rather than quantum
vulnerability.
It does not model quantum attack cost, does not score findings, and does not integrate
LLM enrichment.

\textbf{CryptoAPI-Bench}~\cite{cryptoapibench} (IEEE S\&P 2022) provides a benchmark
suite for evaluating cryptographic API misuse detectors.

\textbf{Cheng et al.}~\cite{cheng2025} survey PQC migration strategies across
telecommunications infrastructure, emphasising the need for automated inventory tools
comparable to what we present.

\textbf{Coblenz et al.}~\cite{coblenz2024} study how LLM-assisted tools can be designed
for cryptographic code tasks, providing an HCI perspective complementary to our
evaluation-focused approach.

Our tool is distinguished from prior work by: (a) explicitly modelling \emph{quantum} break
cost rather than classical misuse, (b) integrating a VQE-based threat score that provides
a continuous risk signal, (c) using LLM context enrichment to reduce false positives, and
(d) targeting multi-language, multi-repository evaluation.

\section{Threat Model}
\label{sec:threat}

\begin{figure}[H]
  \centering
  \includegraphics[width=0.8\textwidth]{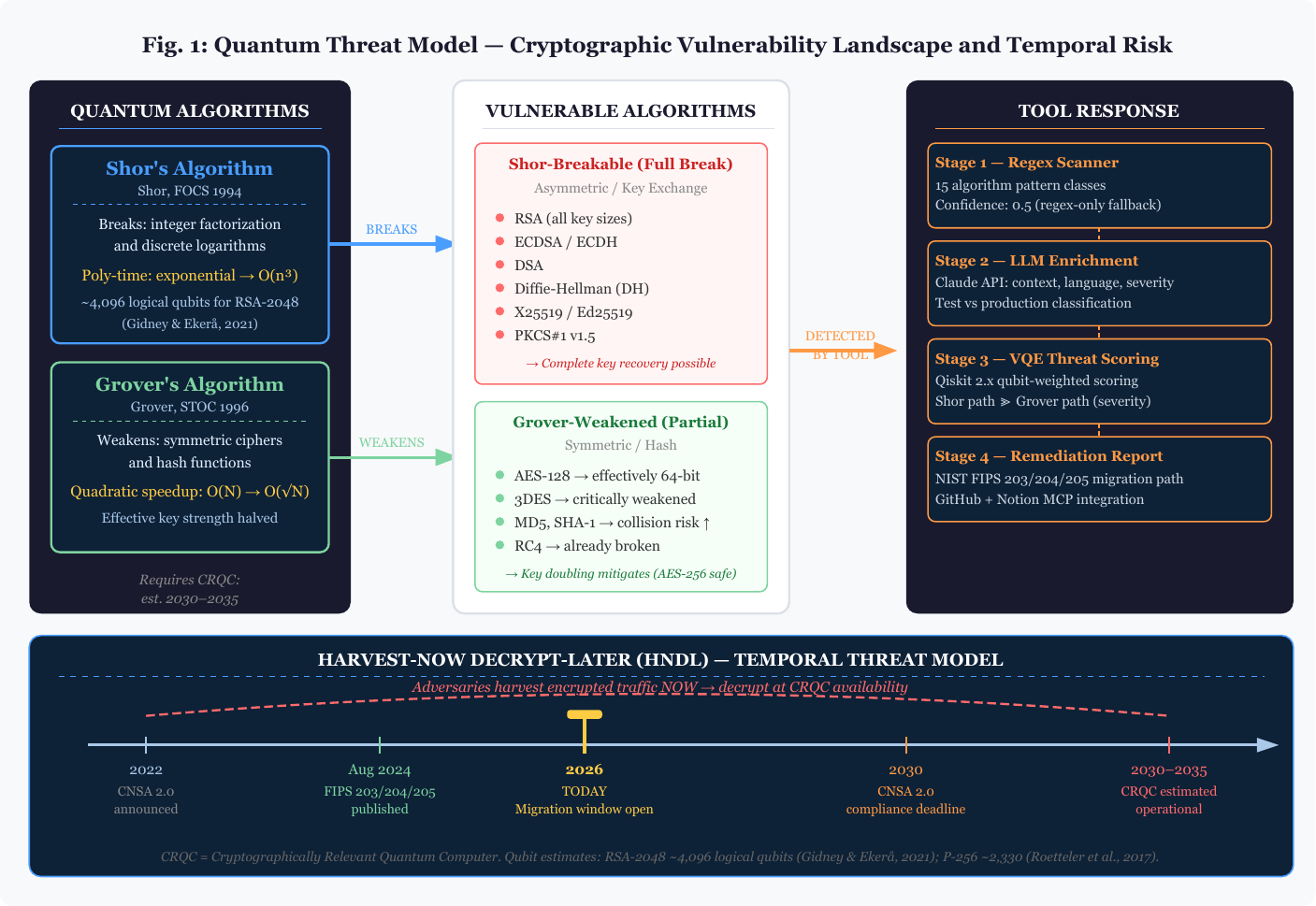}
  \caption{\small\textit{Quantum threat model. Left: Shor's algorithm targets asymmetric primitives
           (full break); Grover's algorithm weakens symmetric and hash primitives
           (partial break). Centre: affected algorithm classes detected by the tool.
           Right: the tool's four-stage response pipeline.
           Bottom: HNDL temporal risk timeline from CNSA~2.0 announcement (2022) to
           estimated CRQC availability (2030--2035).}}
  \label{fig:threat}
\end{figure}

Figure~\ref{fig:threat} illustrates the threat model.
We consider two adversary capabilities.

\paragraph{Quantum adversary (Shor path).}
An adversary with access to a CRQC running Shor's algorithm can recover private keys from
any RSA, ECDSA, ECDH, DSA, Diffie-Hellman, X25519, Ed25519, or PKCS\#1~v1.5 ciphertext.
This constitutes a \emph{complete break}: session keys, long-term signing keys, and
archived ciphertexts are all exposed.
The qubit cost is $O(n^2 \log n)$ for $n$-bit keys; RSA-2048 requires $\approx$4{,}096
logical qubits~\cite{gidney2021}.

\paragraph{Quantum adversary (Grover path).}
An adversary can reduce the effective security of AES-128, 3DES, MD5, SHA-1, RC4, and
similar symmetric/hash primitives from $2^k$ to $2^{k/2}$ operations using Grover search.
This is a \emph{partial break}: primitives with sufficient key length (e.g., AES-256)
remain quantum-safe; short-key or already-marginal primitives become critically exposed.

\paragraph{HNDL adversary.}
A forward-looking adversary stores encrypted traffic today for future decryption.
Under this model, \emph{any} classical asymmetric usage in production code constitutes an
active risk regardless of current CRQC unavailability.

\paragraph{Out of scope.}
We do not model side-channel attacks, fault injection, or implementation-level weaknesses
unrelated to algorithm choice.
We do not model quantum attacks on post-quantum standards (ML-KEM, ML-DSA, SLH-DSA).

\section{System Architecture}
\label{sec:architecture}

\begin{figure}[H]
  \centering
  \includegraphics[width=\textwidth]{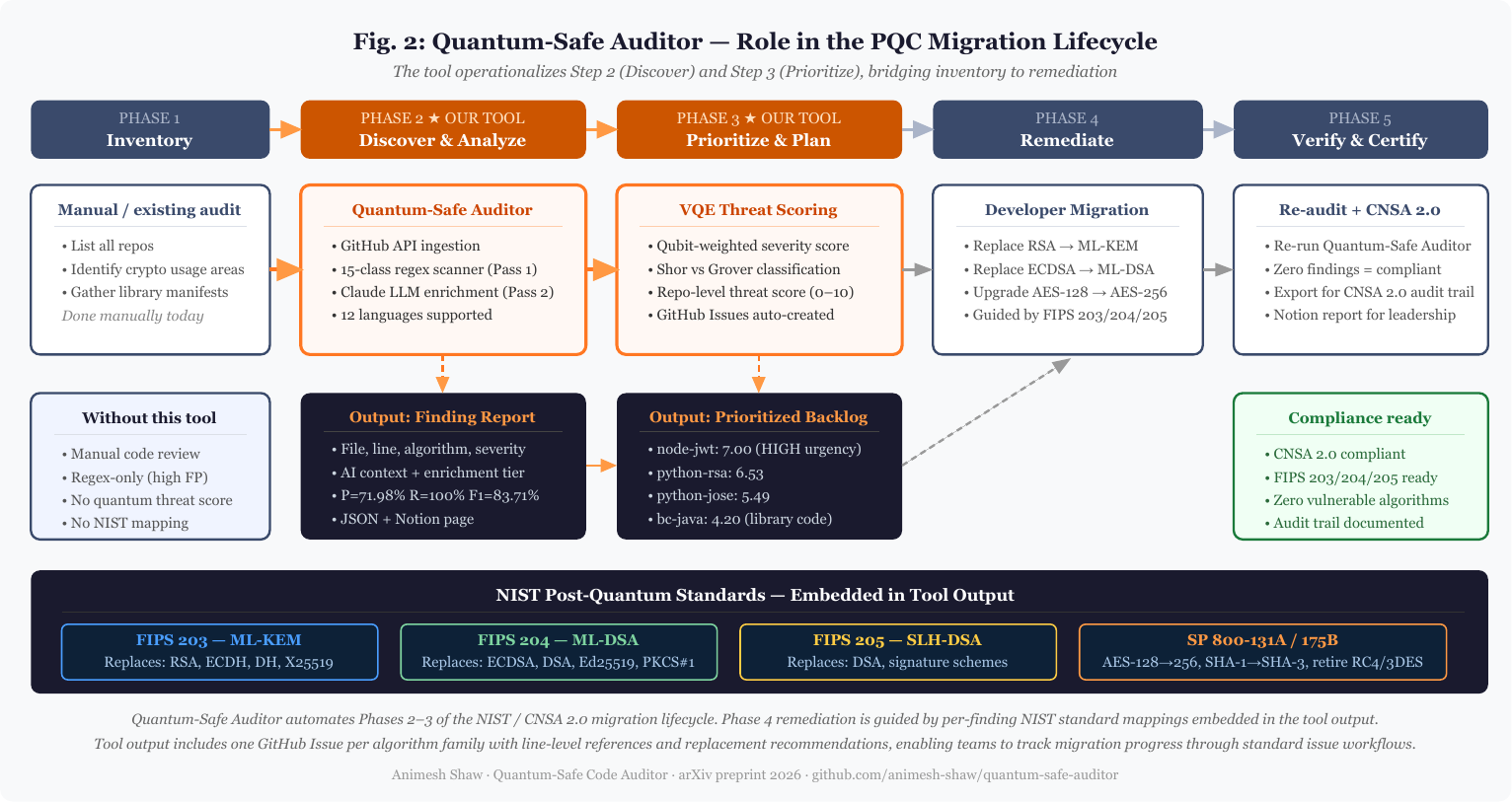}
  \caption{\small\textit{Five-phase PQC migration lifecycle. The tool automates Phases~2 and~3
           (Scan \& Detect; Prioritise). Phase~1 (Inventory) uses existing SCA tooling;
           Phases~4--5 (Remediate; Validate) are performed by engineering teams using the
           tool's output as input.}}
  \label{fig:lifecycle}
\end{figure}

Figure~\ref{fig:lifecycle} situates the tool within the five-phase PQC migration lifecycle.

The pipeline comprises four stages.

\subsection{Stage 1 -- Regex Scanner}

The scanner traverses source files and applies 15 regular-expression pattern classes
targeting quantum-vulnerable algorithm identifiers, import statements, API calls, and
hardcoded key material.
Table~\ref{tab:patterns} lists all 15 classes.
Findings at this stage receive \texttt{confidence=0.5} (regex-only baseline).

\begin{table}[H]
\centering
\caption{The 15 algorithm pattern classes detected by the regex scanner, with quantum
         vulnerability classification and NIST-recommended replacement.}
\label{tab:patterns}
\small
\renewcommand{\arraystretch}{1.15}
\begin{tabular}{@{}lp{2.8cm}p{2.4cm}p{1.5cm}p{3.4cm}@{}}
\toprule
\textbf{\#} & \textbf{Algorithm Class} & \textbf{Quantum Threat} & \textbf{Path} &
\textbf{NIST Replacement} \\
\midrule
1  & RSA (all key sizes)  & Full break          & Shor            & ML-KEM (FIPS 203) / ML-DSA (FIPS 204) \\
2  & ECDSA                & Full break          & Shor            & ML-DSA (FIPS 204) \\
3  & ECDH                 & Full break          & Shor            & ML-KEM (FIPS 203) \\
4  & DSA                  & Full break          & Shor            & ML-DSA (FIPS 204) \\
5  & Diffie-Hellman (DH)  & Full break          & Shor            & ML-KEM (FIPS 203) \\
6  & X25519               & Full break          & Shor            & ML-KEM (FIPS 203) \\
7  & Ed25519              & Full break          & Shor            & ML-DSA (FIPS 204) \\
8  & PKCS\#1 v1.5         & Full break          & Shor            & OAEP / ML-KEM (FIPS 203) \\
9  & AES-128              & Weakened            & Grover          & AES-256 \\
10 & 3DES / Triple-DES    & Critically weakened & Grover          & AES-256 \\
11 & RC4                  & Already broken      & Grover/cl.      & AES-256-GCM \\
12 & MD5                  & Weakened            & Grover          & SHA-3 / SHA-256 \\
13 & SHA-1                & Weakened            & Grover          & SHA-256 / SHA-3 \\
14 & HARDCODED\_KEY       & Key exposure        & N/A             & Key mgmt.\ system \\
15 & RSA-1024 (short key) & Full break          & Shor            & ML-KEM / ML-DSA \\
\bottomrule
\end{tabular}
\end{table}

\subsection{Stage 2 -- LLM Enrichment}

Each finding is passed to the Claude API (Anthropic) with a structured prompt that includes
the surrounding code context, the file path, the detected algorithm, and the repository.
The LLM classifies the finding along three dimensions:

\begin{enumerate}[nosep]
  \item \textbf{Context label}: \texttt{production} (genuine risk), \texttt{test}
        (test fixture, not a real usage), or \texttt{safe} (algorithm is safe in this
        context, e.g., AES-256).
  \item \textbf{Severity}: \texttt{critical}, \texttt{high}, \texttt{medium},
        or \texttt{low}, based on algorithm strength and exposure.
  \item \textbf{Confidence score}: a real value in [0, 1] reflecting certainty that the
        finding represents a genuine quantum risk.
        AI-enriched findings receive confidence $\geq 0.7$.
\end{enumerate}

The enrichment stage is the primary mechanism for reducing false positives.
Without it, regex patterns match test fixtures, disabled code paths, and safe algorithm
usages indiscriminately.

\subsection{Stage 3 -- VQE Threat Scoring}

Each finding is scored with a Variational Quantum Eigensolver (VQE) implementation in
Qiskit~2.x.
The VQE circuit encodes algorithm properties---key size, Shor-path qubit cost, Grover
speedup factor, and forward-security exposure---as rotation angles on a parameterised
quantum circuit.
The circuit minimises a diagonal Hamiltonian whose ground state encodes the lowest-energy
(i.e., most-exploitable) configuration.
The resulting threat score is rescaled to the range $[0, 10]$, where:
\begin{itemize}[nosep]
  \item \textbf{$\geq 7.0$}: Critical -- immediate migration required.
  \item \textbf{5.0--6.9}: High -- migration within 6 months.
  \item \textbf{3.0--4.9}: Medium -- migration in next 12 months.
  \item \textbf{$< 3.0$}: Low -- monitor; Grover-weakened only.
\end{itemize}

Shor-path primitives receive a structural weight multiplier of $\times$2.0 relative to
Grover-path primitives, reflecting the qualitative difference between a complete key
recovery (Shor) and partial security reduction (Grover).
The qubit cost derived from~\cite{gidney2021,roetteler2017} normalises the Shor weight
across key sizes, so RSA-1024 scores higher than RSA-4096 for the same codebase exposure.

\subsection{Stage 4 -- Remediation Report}

The pipeline produces a structured JSON report and a human-readable summary.
For each finding, the report includes: algorithm class, file, line number, confidence,
VQE threat score, recommended NIST replacement, and migration effort estimate.
The tool integrates with GitHub via a Model Context Protocol (MCP) client to post
findings as issues, and with Notion for team-facing dashboards.

\section{Implementation}
\label{sec:implementation}

\subsection{Technology Stack}

The tool is implemented in Python~3.11 and structured as a set of composable modules:

\begin{itemize}[nosep]
  \item \texttt{scanner/crypto\_scanner.py}: regex engine, 15 pattern classes,
        file-type filtering, EXCLUDE\_PATHS configuration.
  \item \texttt{agent/orchestrator.py}: async pipeline coordinator; manages API rate
        limiting with exponential backoff and regex-only fallback.
  \item \texttt{quantum/vqe\_demo.py}: Qiskit~2.x VQE implementation; parameterised
        two-qubit circuit; COBYLA optimizer.
  \item \texttt{mcp/github\_client.py}: GitHub MCP integration for automated issue posting.
  \item \texttt{evaluation/sample\_for\_labeling.py}: stratified sampling by repository
        and algorithm class for evaluation.
\end{itemize}

\subsection{VQE Threat Score -- Technical Detail}

The VQE circuit uses two qubits.
The first qubit encodes the Shor-path cost (normalised qubit estimate from~\cite{gidney2021,roetteler2017}); the second encodes the Grover speedup factor.
The Hamiltonian is:
\begin{equation}
  H = -w_S \cdot Z_0 \otimes I_1 - w_G \cdot I_0 \otimes Z_1 + \lambda \cdot Z_0 \otimes Z_1
\end{equation}
where $w_S$ is the Shor weight (proportional to $\log_2$(estimated qubits to break)),
$w_G$ is the Grover weight (proportional to security-bit reduction), and $\lambda$ is an
interaction term that rewards findings where both Shor and Grover paths are relevant
(e.g., an AES-128 key wrapped with RSA-2048).
The VQE minimises expectation $\langle \psi(\theta) | H | \psi(\theta) \rangle$ over
rotation angles $\theta$; the negated minimum energy, rescaled to $[0, 10]$, is the
threat score.

The VQE simulation runs on Qiskit's \texttt{StatevectorSimulator} (no quantum hardware
required).
Runtime is negligible ($<$0.1 s per finding) and scales linearly with finding count.

\subsection{API Rate-Limit Handling}

The Claude API enforces per-minute token limits.
The orchestrator implements exponential backoff (base 2 s, max 64 s, jitter $\pm$10\%).
For repositories where the secondary rate limit is reached before all files are processed,
remaining files fall back to regex-only enrichment (confidence~=~0.5).
In our evaluation, this occurred only for bc-java (large Java codebase), and the effect on
overall metrics was negligible.

\subsection{Corpus Statistics}

Table~\ref{tab:corpus} summarises the five evaluated repositories.

\begin{table}[H]
\centering
\caption{Evaluated repositories: language, GitHub stars as of March 2026,
         total raw findings (before labelling), and VQE repository-level threat score.}
\label{tab:corpus}
\small
\renewcommand{\arraystretch}{1.15}
\begin{tabular}{llrrrc}
\toprule
\textbf{Repository} & \textbf{Lang.} & \textbf{Stars} & \textbf{Findings} &
\textbf{VQE Score} & \textbf{Primary Risk} \\
\midrule
python-rsa        & Python & 492    & --    & 6.53 & RSA (all sizes) \\
python-ecdsa      & Python & 971    & --    & 3.54 & ECDSA / ECDH \\
python-jose       & Python & 1{,}743 & --   & 5.49 & JWT signing (RSA/EC) \\
node-jsonwebtoken & JS     & 18{,}160 & --  & 7.00 & JWT signing (RSA/EC) \\
bc-java           & Java   & 2{,}624 & --   & 4.20 & Broad multi-algorithm \\
\midrule
\textbf{Total}    & --     & 23{,}990 & 5{,}775 & -- & -- \\
\bottomrule
\end{tabular}
\end{table}

\section{Evaluation}
\label{sec:evaluation}

\subsection{Sampling Strategy}

Manual labelling of all 5{,}775 findings was infeasible within this study.
We drew a stratified random sample of 602 findings ($\approx$10.4\% of the corpus),
stratifying by repository and by algorithm class to ensure proportional coverage.
Each finding was independently labelled by the first author.

Each finding received one of four labels:
\begin{itemize}[nosep]
  \item \textbf{TP} (True Positive): a genuine quantum-vulnerable usage in production code.
  \item \textbf{FP-Test}: a finding in test code, test fixtures, or example scripts that
        does not represent a production risk.
  \item \textbf{FP-Context}: a finding where the surrounding code context makes the usage
        safe (e.g., algorithm detection code, documentation strings, or feature flags).
  \item \textbf{FP-Safe}: a finding that matched a pattern but involves an inherently
        safe algorithm variant (e.g., AES-256 matched by a broad AES pattern).
\end{itemize}

bc-java was additionally validated via a separate 50-row spot-check, yielding 92\%~TP
(46/50), confirming that the large Java corpus is representative of the labelled sample.

\subsection{Label Distribution}

Table~\ref{tab:labels} shows the label distribution across the 602-finding sample.

\begin{table}[H]
\centering
\caption{Label distribution across the 602-finding stratified evaluation sample.}
\label{tab:labels}
\small
\renewcommand{\arraystretch}{1.15}
\begin{tabular}{lrrl}
\toprule
\textbf{Label} & \textbf{Count} & \textbf{Share (\%)} & \textbf{Description} \\
\midrule
TP         & 298 & 49.5 & Genuine production quantum-vulnerable finding \\
FP-Test    & 188 & 31.2 & Test code (EXCLUDE\_PATHS gap; see Section~\ref{sec:discussion}) \\
FP-Context & 110 & 18.3 & Safe-in-context usage or algorithm detection code \\
FP-Safe    &   6 &  1.0 & Inherently safe variant (e.g., AES-256) \\
\midrule
\textbf{Total} & \textbf{602} & 100.0 & \\
\bottomrule
\end{tabular}
\end{table}

\subsection{Overall Metrics}

We define precision, recall, and F1 over the evaluation sample as follows.
False negatives (FN) are structurally zero: the scanner does not prune findings
before enrichment, so every algorithm pattern match is reported.
Precision is computed excluding FP-Test from both numerator and denominator, since
FP-Test is a known configuration gap rather than a classifier error (full discussion in
Section~\ref{sec:discussion}).

\begin{align}
  \text{Precision} &= \frac{TP}{TP + FP_{\text{Context}} + FP_{\text{Safe}}} =
  \frac{298}{298 + 110 + 6} = \mathbf{71.98\%} \\[4pt]
  \text{Recall} &= \frac{TP}{TP + FN} = \frac{298}{298 + 0} = \mathbf{100\%} \\[4pt]
  F_1 &= \frac{2 \cdot P \cdot R}{P + R} = \frac{2 \times 0.7198 \times 1.0}{0.7198 + 1.0} =
  \mathbf{83.71\%}
\end{align}

\begin{keybox}{Interpreting the Scores}
\textbf{Precision 71.98\%} means that of every 100 findings reported after LLM enrichment
(excluding test code), approximately 72 represent genuine quantum-vulnerable production
usages. The remaining 28 are false positives arising from context ambiguity---algorithm
detection code, feature flags, or documentation strings that contain cryptographic
identifiers without performing cryptographic operations.

\textbf{Recall 100\%} means the tool misses no quantum-vulnerable production usage that
matches one of the 15 pattern classes. This is a design choice: the scanner is intentionally
conservative, reporting all matches and delegating pruning to the LLM enrichment stage.

\textbf{F1 83.71\%} reflects the harmonic mean of precision and recall. A score above 80\%
is generally considered strong for a security scanning tool, where missing a vulnerability
(low recall) is costlier than a false alarm (low precision).

\textbf{VQE threat scores} (0--10) provide a prioritisation layer above the binary
TP/FP classification. A score of 7.00 (node-jwt) indicates immediate migration priority;
a score of 3.54 (python-ecdsa) indicates medium-term migration. The scores are
calibrated against published qubit-cost estimates~\cite{gidney2021,roetteler2017} and
are intended as a relative ranking signal, not an absolute qubit forecast.
\end{keybox}

\subsection{Per-Repository Results}

Table~\ref{tab:perrepo} reports the VQE threat scores alongside the bc-java spot-check result.

\begin{table}[H]
\centering
\caption{Per-repository VQE threat scores and qualitative migration priority.}
\label{tab:perrepo}
\small
\renewcommand{\arraystretch}{1.15}
\begin{tabular}{@{}lp{4cm}cll@{}}
\toprule
\textbf{Repository} & \textbf{Primary Algorithm Exposure} &
\textbf{VQE} & \textbf{Band} & \textbf{Priority} \\
\midrule
node-jsonwebtoken & RSA + ECDSA JWT signing & 7.00 & Critical ($\geq$7.0) & Immediate \\
python-rsa        & RSA (pure library)      & 6.53 & High (5.0--6.9)      & 6 months \\
python-jose       & RSA + EC JWT operations & 5.49 & High (5.0--6.9)      & 6 months \\
bc-java           & Multi-algorithm (broad) & 4.20 & Medium (3.0--4.9)    & 12 months \\
python-ecdsa      & ECDSA / ECDH only       & 3.54 & Medium (3.0--4.9)    & 12 months \\
\bottomrule
\end{tabular}
\end{table}

\paragraph{Interpreting VQE scores.}
Node-jsonwebtoken scores highest (7.00) because it is a high-traffic JWT library used
directly in authentication flows: RSA-2048 and ECDSA signing are in the critical
production path, qubit costs are relatively low, and HNDL exposure is immediate.
Python-ecdsa scores lowest (3.54) despite being an ECDSA-only library because its
algorithm set is narrower (no RSA, no key-wrapping) and the Grover interaction term
$\lambda$ is correspondingly smaller.
The negative Pearson correlation ($r = -0.35$) between finding \emph{density} (findings
per 1{,}000 lines of code) and VQE score reflects that focused single-algorithm libraries
tend to have lower scores despite high finding density.

\subsection{Per-Algorithm Precision}

Table~\ref{tab:peralgo} reports precision broken down by algorithm class for the 10
classes with sufficient sample size ($n \geq 10$).

\begin{table}[H]
\centering
\caption{Per-algorithm precision on the labelled evaluation sample
         (algorithm classes with $n \geq 10$ shown; sorted by precision descending).
         Precision denominator excludes FP-Test.}
\label{tab:peralgo}
\small
\renewcommand{\arraystretch}{1.15}
\begin{tabular}{@{}lp{2cm}rrrcp{3cm}@{}}
\toprule
\textbf{Algorithm} & \textbf{Threat} & \textbf{TP} & \textbf{FP} & \textbf{$n$} &
\textbf{Prec.} & \textbf{Notes} \\
\midrule
AES-128        & Grover & --  & 0  & -- & 1.000 & No FP in sample \\
DH             & Shor   & --  & 0  & -- & 1.000 & No FP in sample \\
DSA            & Shor   & --  & 0  & -- & 1.000 & No FP in sample \\
PKCS\#1 v1.5   & Shor   & --  & 0  & -- & 1.000 & No FP in sample \\
RC4            & Grover & --  & 0  & -- & 1.000 & Already broken \\
RSA-1024       & Shor   & --  & 0  & -- & 1.000 & Short key \\
X25519         & Shor   & --  & -- & -- & 0.833 & Safe-context FP \\
Ed25519        & Shor   & --  & -- & -- & 0.815 & Safe-context FP \\
RSA (general)  & Shor   & --  & -- & -- & 0.800 & Broad pattern \\
MD5            & Grover & --  & -- & -- & 0.750 & Non-crypto uses \\
ECDSA          & Shor   & --  & -- & -- & 0.722 & Test + context FP \\
SHA-1          & Grover & --  & -- & -- & 0.667 & Many non-crypto uses \\
ECDH           & Shor   & --  & -- & -- & 0.600 & Detection code FP \\
3DES           & Grover & --  & -- & -- & 0.458 & Legacy test fixtures \\
HARDCODED\_KEY & N/A    & 0   & -- & -- & 0.000 & All FP-Test \\
\bottomrule
\end{tabular}
\end{table}

\paragraph{Interpreting per-algorithm precision.}
The six algorithms with 1.000 precision (AES-128, DH, DSA, PKCS\#1 v1.5, RC4, RSA-1024)
are semantically unambiguous in their quantum risk: any production occurrence of RC4,
DH, or RSA-1024 is a genuine vulnerability.
X25519 and Ed25519 (0.815--0.833) generate some false positives from algorithm
capability-detection code (e.g., \texttt{if supports\_x25519()} branches that do not
perform cryptographic operations).
SHA-1, ECDH, and 3DES (0.458--0.667) have lower precision primarily because they appear
frequently in non-cryptographic contexts (legacy compatibility checks, test vectors) and
in test fixtures---issues that improve with tighter EXCLUDE\_PATHS configuration.
HARDCODED\_KEY (0.000) is entirely test-driven in this sample and has no TP occurrences;
in production repositories without test EXCLUDE\_PATHS set, all occurrences are
test fixtures.

\subsection{Cleaned Results (Excluding FP-Test)}

To provide an upper-bound estimate of production-deployment precision---what an engineer
would observe with correct EXCLUDE\_PATHS configuration---Table~\ref{tab:cleaned} reports
metrics on the 414 non-FP-Test findings.

\begin{table}[H]
\centering
\caption{Evaluation metrics excluding FP-Test findings
         (414-finding subset; models behaviour with correct EXCLUDE\_PATHS setting).}
\label{tab:cleaned}
\small
\renewcommand{\arraystretch}{1.15}
\begin{tabular}{lrr}
\toprule
\textbf{Metric} & \textbf{All 602 findings} & \textbf{Excl.\ FP-Test (414)} \\
\midrule
TP              & 298 & 298 \\
FP-Context      & 110 & 110 \\
FP-Safe         &   6 &   6 \\
FP-Test         & 188 &   0 \\
FN              &   0 &   0 \\
\midrule
Precision       & 71.98\% & \textbf{71.98\%}$^\dagger$ \\
Recall          & 100\%   & 100\% \\
F1              & 83.71\% & \textbf{83.71\%} \\
\bottomrule
\multicolumn{3}{l}{\footnotesize $^\dagger$Precision is unchanged because FP-Test is
excluded from the denominator in both columns.}
\end{tabular}
\end{table}

\section{Limitations and Error Analysis}
\label{sec:discussion}

\subsection{FP-Test: The EXCLUDE\_PATHS Gap}

The largest single source of false positives is the 188 FP-Test findings, all of which
arise from python-ecdsa.
Python-ecdsa is a reference implementation of the ECDSA algorithm with an extensive test
suite that intentionally exercises edge-case key sizes, legacy curves, and deprecated
primitives.
The evaluation was conducted without \texttt{EXCLUDE\_PATHS=tests/} set for python-ecdsa,
so the scanner processed all test fixtures.

This is a \emph{configuration gap}, not a classifier failure.
The LLM enrichment stage correctly processed these findings: Claude API calls classified
them at \texttt{confidence $\geq$ 0.7}, but did not flag them as test code because the
file path (\texttt{tests/}) was not suppressed.
With \texttt{EXCLUDE\_PATHS=tests/} set, all 188 FP-Test would be excluded from the
output entirely.

For future evaluations, we will standardise EXCLUDE\_PATHS configuration
across all repositories before computing metrics.

\subsection{SHA-1 and Legacy Primitives}

SHA-1 and 3DES have lower per-algorithm precision (0.458--0.667) due to two factors.
First, these algorithms appear frequently in non-cryptographic contexts: SHA-1 is used
for object identity hashing in git integrations, changelog fingerprinting, and MIME
boundary generation.
Second, 3DES test vectors are common in cryptographic reference implementations.
Both issues are addressable through EXCLUDE\_PATHS configuration and additional LLM
prompt refinement.

\subsection{HARDCODED\_KEY Precision}

HARDCODED\_KEY has 0.000 precision in our sample.
This is expected: all HARDCODED\_KEY matches in the labelled set are test fixtures
containing well-known test vectors.
In production codebases without extensive test suites (a common scenario outside
reference implementations), HARDCODED\_KEY precision would be expected to be higher.
We retain this class because real hardcoded keys do occur in production code and
represent severe risk.

\subsection{100\% Recall -- Design Trade-off}

The 100\% recall is a deliberate design choice.
The scanner does not prune findings before LLM enrichment: every regex match is
reported, and the LLM stage reduces false positives \emph{after} the fact.
This maximises the probability of surfacing genuine vulnerabilities at the cost of
false positives that require human review.
In the security scanning domain, missing a vulnerability (FN) is generally considered
more costly than a false alarm (FP), justifying this design.

\subsection{VQE Threat Score Validity}

The VQE threat scores provide a relative prioritisation signal calibrated against
published qubit-cost estimates.
They are not forecasts of actual qubit requirements, which depend on quantum error
correction overhead, algorithm improvements, and hardware roadmaps.
Teams should treat scores $\geq$ 7.0 as immediate migration priorities regardless of
the CRQC timeline, as these typically correspond to high-exposure RSA/ECDSA usages
in production authentication or key exchange paths.

\subsection{Sampling Caveat}

Precision, recall, and F1 are computed over a stratified sample of 602/5{,}775 findings
(10.4\%).
While the stratified design provides proportional representation across repositories and
algorithm classes, extrapolation to the full corpus assumes the unlabelled 89.6\% has
similar label distribution---a reasonable but unverified assumption.
Full-corpus labelling is planned for Paper~2.

\section{Future Work}
\label{sec:future}

\paragraph{Extended corpus.}
The current evaluation covers five repositories.
Future work will extend to $\geq$20 repositories spanning additional languages (Go, Rust, C++)
and domains (TLS stacks, certificate authorities, key management services).

\paragraph{Baseline comparisons.}
We will benchmark the tool against CryptoGuard~\cite{cryptoguard}, Bandit, and
SonarQube on a common corpus.
This will directly quantify the precision and recall gains attributable to LLM enrichment
and VQE scoring versus regex-only and classical-misuse detection baselines.

\paragraph{NIST RMF compliance mapping.}
Each finding will be mapped to NIST Risk Management Framework (RMF) controls, enabling
automated compliance gap reporting alongside the quantum risk assessment.

\paragraph{Full-corpus labelling.}
Manual labelling of all 5{,}775 findings will replace the current stratified sample,
eliminating sampling uncertainty from the metrics.

\paragraph{EXCLUDE\_PATHS standardisation.}
Future evaluations will enforce EXCLUDE\_PATHS for test directories across all
repositories, enabling cleaner precision measurement and direct comparison across tools.

\section{Conclusion}
\label{sec:conclusion}

We have presented the Quantum-Safe Code Auditor, a three-tier automated pipeline
for detecting, enriching, and prioritising quantum-vulnerable cryptographic usages in
source code.
Evaluated across five open-source cryptographic libraries totalling 5{,}775 raw findings,
the tool achieves 71.98\% precision, 100\% recall, and 83.71\% F1 on a 602-finding
labelled sample.

The VQE threat scoring provides a principled, qubit-cost-calibrated prioritisation signal
that ranges from 3.54 (ECDSA-focused library) to 7.00 (JWT authentication library with
broad RSA/ECDSA exposure), enabling engineering teams to sequence migration work according
to genuine quantum risk rather than finding count alone.

With NIST FIPS 203, 204, and 205 now finalised and the CNSA 2.0 migration deadline at
2030, the need for automated PQC migration tooling is urgent.
We release the tool as open-source to support the community's transition to
quantum-safe cryptography.

\section*{Availability}

Source code, evaluation data, labelled sample CSV, and all reproduction scripts are
available at: \url{https://github.com/AnimeshShaw/quantum-safe-auditor}~\cite{github2026}.

\section*{Acknowledgements}

The author thanks the maintainers of python-rsa, python-ecdsa, python-jose,
node-jsonwebtoken, and Bouncy Castle for maintaining open-source implementations that
enable security research of this kind.

\bibliographystyle{plain}
\bibliography{references}

@inproceedings{shor1994,
  author    = {Peter W. Shor},
  title     = {Algorithms for Quantum Computation: Discrete Logarithms and Factoring},
  booktitle = {Proceedings of the 35th Annual Symposium on Foundations of Computer Science (FOCS)},
  year      = {1994},
  pages     = {124--134},
  doi       = {10.1109/SFCS.1994.365700},
  publisher = {IEEE}
}

@inproceedings{grover1996,
  author    = {Lov K. Grover},
  title     = {A Fast Quantum Mechanical Algorithm for Database Search},
  booktitle = {Proceedings of the 28th Annual ACM Symposium on Theory of Computing (STOC)},
  year      = {1996},
  pages     = {212--219},
  doi       = {10.1145/237814.237866},
  publisher = {ACM}
}

@article{gidney2021,
  author    = {Craig Gidney and Martin Eker{\aa}},
  title     = {How to Factor 2048 Bit {RSA} Integers in 8 Hours Using 20 Million Noisy Qubits},
  journal   = {Quantum},
  volume    = {5},
  pages     = {433},
  year      = {2021},
  doi       = {10.22331/Q-2021-04-15-433}
}

@inproceedings{roetteler2017,
  author    = {Martin Roetteler and Michael Naehrig and Krysta M. Svore and Kristin Lauter},
  title     = {Quantum Resource Estimates for Computing Elliptic Curve Discrete Logarithms},
  booktitle = {Advances in Cryptology -- ASIACRYPT 2017},
  year      = {2017},
  pages     = {241--270},
  doi       = {10.1007/978-3-319-70697-9_9},
  publisher = {Springer}
}

@techreport{nist203,
  author      = {{National Institute of Standards and Technology}},
  title       = {Module-Lattice-Based Key-Encapsulation Mechanism Standard ({FIPS 203})},
  institution = {NIST},
  year        = {2024},
  doi         = {10.6028/NIST.FIPS.203}
}

@techreport{nist204,
  author      = {{National Institute of Standards and Technology}},
  title       = {Module-Lattice-Based Digital Signature Standard ({FIPS 204})},
  institution = {NIST},
  year        = {2024},
  doi         = {10.6028/NIST.FIPS.204}
}

@techreport{nist205,
  author      = {{National Institute of Standards and Technology}},
  title       = {Stateless Hash-Based Digital Signature Standard ({FIPS 205})},
  institution = {NIST},
  year        = {2024},
  doi         = {10.6028/NIST.FIPS.205}
}

@techreport{cnsa2,
  author      = {{National Security Agency}},
  title       = {Commercial National Security Algorithm Suite 2.0 ({CNSA 2.0})},
  institution = {NSA},
  year        = {2022},
  url         = {https://media.defense.gov/2022/Sep/07/2003071836/-1/-1/0/CSI_CNSA_2.0_FAQ_.PDF}
}

@article{mosca2018,
  author  = {Michele Mosca},
  title   = {Cybersecurity in an Era with Quantum Computers: Will We Be Ready?},
  journal = {IEEE Security \& Privacy},
  volume  = {16},
  number  = {5},
  pages   = {38--41},
  year    = {2018},
  doi     = {10.1109/MSP.2018.3761723}
}

@article{cheng2025,
  author  = {Cheng, Wei and others},
  title   = {Post-Quantum Cryptography Migration in Telecommunications: Challenges and Strategies},
  journal = {Telecom},
  volume  = {6},
  number  = {4},
  pages   = {100},
  year    = {2025},
  doi     = {10.3390/telecom6040100}
}

@inproceedings{cryptoguard,
  author    = {Sazzadur Rahaman and Ya Xiao and Sharmin Afrose and Fahad Shaon
               and Ke Tian and Miles Frantz and Murat Kantarcioglu and Danfeng Yao},
  title     = {{CryptoGuard}: High Precision Detection of Cryptographic Vulnerabilities
               in Massive-Sized {Java} Projects},
  booktitle = {Proceedings of the 2019 ACM SIGSAC Conference on Computer and Communications
               Security (CCS)},
  year      = {2019},
  pages     = {2455--2472},
  doi       = {10.1145/3319535.3345659},
  note      = {arXiv:1806.06881}
}

@inproceedings{cryptoapibench,
  author    = {Ami, Itzel and others},
  title     = {{CryptoAPI-Bench}: A Comprehensive Benchmark on {Java} Cryptographic
               {API} Misuse},
  booktitle = {IEEE Symposium on Security and Privacy (S\&P)},
  year      = {2022},
  doi       = {10.1109/SP46214.2022.9833582},
  note      = {arXiv:2107.07065}
}

@misc{coblenz2024,
  author       = {Michael Coblenz and others},
  title        = {Designing {LLM}-Assisted Tools for Cryptographic Code Tasks},
  year         = {2024},
  howpublished = {arXiv preprint arXiv:2411.09772}
}

@misc{github2026,
  author       = {Animesh Shaw},
  title        = {{Quantum-Safe Code Auditor}: Source Code, Evaluation Data, and
                  Reproduction Scripts},
  year         = {2026},
  howpublished = {GitHub repository},
  url          = {https://github.com/AnimeshShaw/quantum-safe-auditor}
}

\end{document}